\documentclass[aps,floats,twocolumn,epsf,prl,showpacs]{revtex4} 
\usepackage{amssymb} 
\usepackage{graphicx} 
\usepackage{amsmath} 
\usepackage{subfigure} 
 
\begin{document}
 
\title{Turning a nickelate Fermi surface into a cuprate-like one through 
heterostructuring} 
\author{ P. Hansmann$^{1,2}$, Xiaoping Yang$^1$, A. Toschi$^{1,2}$, G. 
Khaliullin$^1$, O. K. Andersen$^1$, K. Held$^2$} 
\affiliation{$^1$ Max-Planck-Institut f\"ur Festk\"orperforschung, Heisenbergstrasse 1, 
D-70569 Stuttgart, Germany \\ 
$^2$ Institute for Solid State Physics, Vienna University of Technology, 
1040 Vienna, Austria} 
\date{Version 2, \today} 
 
\begin{abstract} 
Using the local density approximation and its combination with dynamical 
mean-field theory, we show that electronic correlations induce a 
single-sheet, cuprate-like Fermi surface for hole-doped 1/1 LaNiO$_{3}$/LaAlO%
$_{3}$ heterostructures, even though both $e_{g}$ orbitals contribute to it. 
The Ni $3d_{3z^{2}-1}$ orbital plays the role of the axial Cu $4s$-like 
orbital in the cuprates. These two results indicate that "orbital 
engineering"\ by means of heterostructuring should be possible. As we 
also find strong antiferromagnetic correlations, the low-energy electronic and 
spin excitations in nickelate heterostructures resemble those of 
high-temperature cuprate superconductors. 
\end{abstract} 
 
\maketitle 
 
\let\n=\nu \let\o =\omega \let\s=\sigma 
 
The discovery of high-temperature superconductivity (HTSC) in hole-doped 
cuprates \cite{Bednorz86} initiated the quest for finding related 
transition-metal oxides with comparable or even higher transition 
temperatures. In some systems such as ruthenates \cite{Maeno94} and 
cobaltates \cite{Takada03a} superconductivity has been found. However, in 
these $t_{2g}$ systems superconductivity is very different from that in 
cuprates and transition temperatures ($T_{c}$'s) are considerably lower. 
 
As it became possible to grow transition-metal oxides in heterostructures, 
this quest got a new direction: Novel effectively two-dimensional (2D) 
systems could be engineered. But which oxides, besides cuprates, are most 
promising for getting high $T_{c}$'s? 
 
The basic band structure of the hole-doped cuprates is that of a single 2D 
Cu 3$d_{x^{2}-y^{2}}$-like band which is less than half-filled 
(configuration $d^{9-h}$). In this situation, antiferromagnetic fluctuations 
prevail and are often believed to mediate the superconductivity. The Fermi 
surface (FS) from this $x^{2}-y^{2}$ band has been observed in many 
overdoped cuprates and found to agree with the predictions of 
density-functional (LDA) band theory. 
 
Recently the following idea for arriving at a cuprate-like situation in 
nickelates was presented \cite{Chaloupka}: Bulk LaNiO$_{3}$ $\left( 
d^{7}\right) $ has one electron in two degenerate $e_{g}$ bands, but 
sandwiching a LaNiO$_{3}$ layer between layers of an insulating oxide such 
as LaAlO$_{3}$ will confine the $3z^{2}-1$ orbital in the $z$-direction and 
may remove this band from the Fermi level, thus leaving the electron in the $%
x^{2}-y^{2}$ band. The possibility of finding bulk nickelates with an 
electronic structure analogous to that of cuprates was discarded a while ago  
\cite{Anisimov99}, but heterostructures offer new perspectives. 
 
Indeed, a major reconstruction of orbital states at oxide interfaces may 
recently have been observed \cite{Chakh07}, and this kind of phenomenon 
could lead to novel phases not present in the bulk. Extensive theoretical 
studies of mechanisms for orbital selection in correlated systems \cite{OSMT} 
have revealed the complexity of this problem, where details of the 
electronic structure and lattice distortions play decisive roles. It is 
therefore crucial to examine nickelate heterostructures by means of 
state-of-the-art theoretical methods and find the optimal conditions for $%
x^{2}-y^{2}$ orbital selection. 
 
In this Letter we present results of electronic-structure calculations using 
the merger \cite{LDADMFT} of density-functional (LDA) band theory, which 
provides an \emph{ab initio} description of the materials chemistry, and the 
dynamical mean-field theory (DMFT) \cite{DMFT}, which includes electronic 
correlations. We find that the hopping between the $x^{2}-y^{2}$ and $%
3z^{2}-1$ orbitals substantially reduces the effects of correlations in the $%
3z^{2}-1$ orbital. In this respect, $e_{g}$ electrons behave very 
differently than the $t_{2g}$ electrons, which have \emph{no} inter-orbital 
hopping on a square lattice. Nevertheless, we do find that the correlations 
may sufficiently shift the bottom of the hybridizing $e_{g}$ bands 
relatively to each other to yield a FS with only \emph{one} sheet. This 
sheet has predominantly $x^{2}-y^{2}$ character and a shape like in the 
cuprates with the highest $T_{c\,\max }$ ($T_{c}$ at optimum hole doping)  
\cite{Pavarini01}, but even more extreme. Moreover, stretching the in-plane 
lattice constants by suitable choice of substrate reduces the 
correlation-strength needed to produce a single-sheet FS. Since we also find 
strong antiferromagnetic fluctuations, somewhat larger than in the cuprates, 
nickelate heterostructures hold the basic ingredients for high-temperature 
superconductivity. 
 
\begin{figure*}[tb] 
\centering   \includegraphics[width=4.5cm, angle=-90]{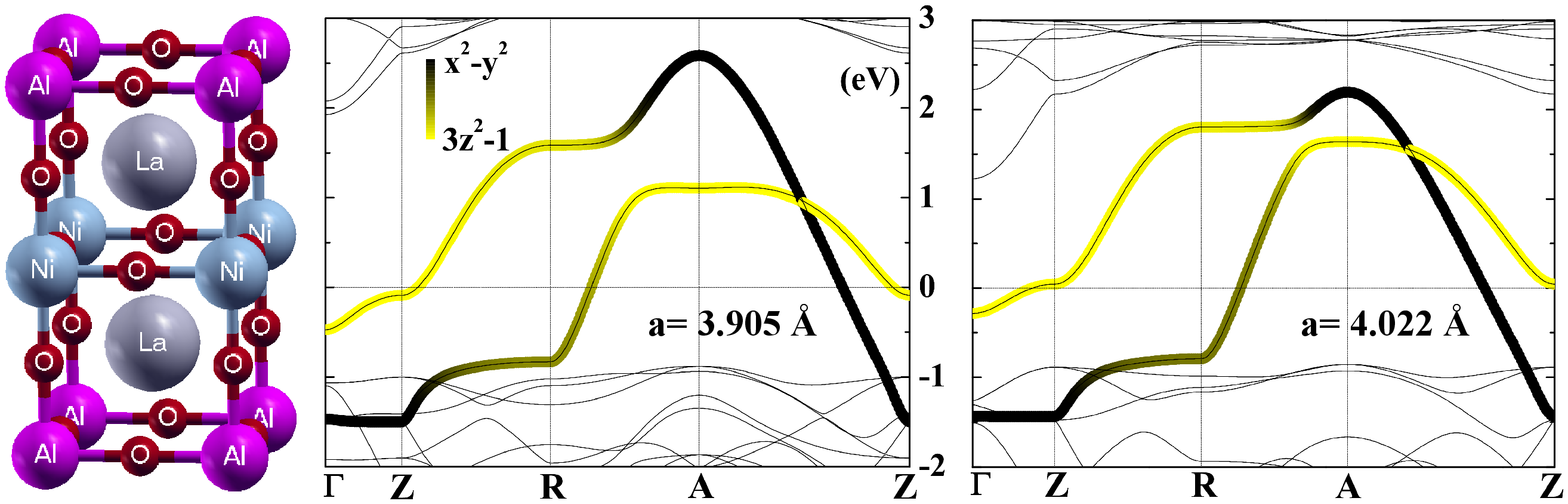}  
\caption{The 1/1 LaNiO$_{3}$/LaAlO$_{3}$ heterostructure (left) and its LDA (%
$N$MTO) bandstructures without (center) and with (right) strain. The Bloch 
vector is along the lines $\Gamma \left( 0,0,0\right) $ $-$ $\mathrm{Z}%
\left( 0,0,\frac{\protect\pi }{c}\right) $ $-$ $\mathrm{R}\left( 0,\frac{%
\protect\pi }{a},\frac{\protect\pi }{c}\right) $ $-$ $\mathrm{A}\left( \frac{%
\protect\pi }{a},\frac{\protect\pi }{a},\frac{\protect\pi }{c}\right) $ $-$ $%
\mathrm{Z}\left( 0,0,\frac{\protect\pi }{c}\right) .$ The shading gives the $%
x^{2}-y^{2}$ vs $3z^{2}-1$ $e_{g}$ Wannier-function character. } 
\label{figure1} 
\end{figure*} 
 
Here we give results for the simplest, 1/1 superlattice LaNiO$_{3}$/LaAlO$%
_{3}\,$=$\,$LaO-NiO$_{2}$-LaO-AlO$_{2}$ shown in the left-hand side of Fig.$%
\,$1. For the in-plane lattice constant $a$ we first took that of SrTiO$_{3}, 
$ often used as substrate, whereby the Ni-O and Al-O distance in the $x$- 
and $y$-directions became: $x_{\mathrm{Ni}\text{\textrm{-}}\mathrm{O}}=1.95\, 
$\AA , not far from the value\ in pseudo-cubic LaNiO$_{3}$. The lattice 
constant $c$ we took as the sum of those of pseudo-cubic LaNiO$_{3}$ and 
LaAlO$_{3},$ whereafter the position of apical O was relaxed within the LDA  
\cite{VASP} to yield: $z_{\mathrm{Ni}\text{-}\mathrm{O}}=1.91\,\mathrm{%
\mathring{A},}$ i.e. 2\% smaller than $x_{\mathrm{Ni}\text{-}\mathrm{O}}.$ 
Next, we expanded the LaNiO$_{3}$/LaAlO$_{3}$ heterostructure in the $x$-and  
$y$-directions by 3\%, as might be achieved by growing LaNiO$_{3}$/LaAlO$_{3} 
$ on a PrScO$_{3}$ substrate, to yield $x_{\mathrm{Ni}\text{\textrm{-}}%
\mathrm{O}}=2.01\,$\AA . With the concomitant 6\% contraction in the $z$%
-direction, relaxation of the apical-oxygen position within the LDA finally 
lead to: $z_{\mathrm{Ni}\text{-}\mathrm{O}}=1.81\,\mathrm{\mathring{A}.}$ 
 
Fig. 1 shows the LDA energy bands for the two differently strained 
heterostructures in a 5-eV region around the $d^{7}$ Fermi level $\left(  
\mathrm{\equiv }0\right) .$ The two solid bands are the 1/4-full Ni-O $%
pd\sigma $ antibonding $e_{g}$ bands, which are pushed up above the less 
antibonding Ni-O $pd\pi $ $t_{2g}$ bands (thin bands) lying below $-1$ eV 
and well above the Ni-O, Al-O, and La-O bonding bands below the frame of the 
figure. The antibonding Al-O and La-O bands (thin bands above 1-2 eV) lie 
respectively 9 and $\sim 5$ eV above their bonding counterparts, and as a 
result there is a comfortable 2-3 eV gap above the top of the antibonding $%
t_{2g}$ bands in which the two antibonding $e_{g}$ bands reside. 
 
The shading (coloring) of the $e_{g}$ bands gives the relative $x^{2}-y^{2}$ 
and $3z^{2}-1$ characters in the Wannier-function representation of these 
two bands, as calculated with the $N$th-order muffin-tin-orbital ($N$MTO) 
method and $N\mathrm{=}2$ \cite{NMTO}. We see that in the "nodal" $k_{x}%
\mathrm{=}k_{y}$ plane containing the $\Gamma \mathrm{Z}$ and $\mathrm{AZ}$ 
lines the $x^{2}-y^{2}$ $\left( \left\vert m\right\vert \mathrm{=}2\right) $ 
and $3z^{2}-1$ $\left( m\mathrm{=}0\right) $ Wannier orbitals cannot mix  
\cite{ttprime}. The bottoms of both bands are along $\Gamma \mathrm{Z,}$ 
i.e. for $k_{x}\mathrm{=}k_{y}\mathrm{=}0\mathrm{.}$ That of the $x^{2}-y^{2} 
$ band is at $-1.5\,\mathrm{eV}$\emph{\ }and does not disperse with $k_{z},$ 
while that of the $3z^{2}-1$ band is at $-0.5$ eV at $\Gamma $ and disperses 
upwards to $-0.1$ eV at $\mathrm{Z}$. The bottom of the $3z^{2}-1$ band is 
thus $1$ eV $\approx 1/4$ $e_{g}$ bandwidth above that of the $x^{2}-y^{2}$ 
band. Straining by 3\% is seen to shift the bottom of the $3z^{2}-1$ band up 
by further 0.2 eV. The LDA FS thus has \emph{two} sheets, and reducing it to 
one would require moving the $3z^{2}-1$ band above the $x^{2}-y^{2}$ band at  
$\Gamma $ by an additional $0.5$~eV for the unstrained and by an additional 
0.3 eV for the strained superlattice. 
 
That the $x^{2}-y^{2}$ Wannier orbital is more populated than $3z^{2}-1$ 
(the ratio is 70/30 for the unstrained superlattice) is mainly due to the  
\emph{confinement} in the $z$-direction. Consider for simplicity the 
dispersions in the $k_{x}\mathrm{=}\pm k_{y}\mathrm{\equiv }k$ planes where 
the $3z^{2}-1$ and $x^{2}-y^{2}$ orbitals do not hybridize: In cubic, bulk 
LaNiO$_{3}$, $\varepsilon _{3z^{2}-1}\left( k,k_{z}\right) \approx -\cos 
ak-2\cos ak_{z},$ with respect to the center of the $e_{g}$ band and in 
units of $\left\vert t_{dd\sigma }\right\vert $, while $\varepsilon 
_{x^{2}-y^{2}}\left( k,k_{z}\right) \approx -3\cos ak$ independently of $%
k_{z}$ because $t_{dd\delta }$ is negligible. This means that both bands 
extend from $-3\left\vert t_{dd\sigma }\right\vert $ at $\left( 0,0,0\right)  
$ to $+3\left\vert t_{dd\sigma }\right\vert $ at $\left( \frac{\pi }{a},%
\frac{\pi }{a},\frac{\pi }{a}\right) $ in the bulk. Substituting now every 
second LaNiO$_{3}$ layer by an "insulating" LaAlO$_{3}$ layer, forces the 
Bloch waves to have nodes approximately at the AlO$_{2}$ planes, so that 
only waves with $k_{z}\gtrsim \frac{\pi }{2a}\sim \frac{\pi }{c}$ are 
allowed. As a consequence, the bottom of the $3z^{2}-1$ band is pushed up by  
$\mathrm{\sim }2\left\vert t_{dd\sigma }\right\vert ,$ i.e. by $\mathrm{\sim  
}1/3$ the $e_{g}$ bandwidth. The exact position of the nodes, and hence the 
upwards shift of the $3z^{2}-1$ band, depends on the scattering properties 
of the insulating layer. This suggests that the band structure can be tuned 
by choice of the insulating layer. 
 
A further factor influencing the orbital separation is the tetragonal 
Jahn-Teller (JT) distortion of the nickel-centered oxygen octahedron. Since 
the $x^{2}-y^{2}$ and $3z^{2}-1$ Wannier orbitals antibond with oxygen, 
flattening the octahedron $\left( z_{\mathrm{Ni}\text{-}\mathrm{O}}<x_{%
\mathrm{Ni}\text{-}\mathrm{O}}\right) $ moves the energy of the former 
orbital down, and that of the latter up. However, this crystal-field 
splitting is little effective in achieving orbital separation for 
configuration $d^{7-h}$ because the $e_{g}$ Bloch sums at the \emph{bottom} 
of the cubic band at $\left( 0,0,0\right) $ have \emph{no} oxygen character, 
so only energies higher up in the $e_{g}$ band are effected. For 
JT-flattening to be effective, confinement is therefore a prerequisite. This 
is clearly seen from the LDA bands for the 3\% strained superlattice on the 
right-hand side of Fig. 1: Whereas the strain moves the top of the $%
x^{2}-y^{2}$ band down and that of the $3z^{2}-1$ band up, the bottom of the  
$x^{2}-y^{2}$ band is \emph{not} affected, and that of the $3z^{2}-1$ band 
is shifted up only because it has antibonding oxygen character corresponding 
to $k_{z}\mathrm{\sim }\frac{\pi }{2a}$ rather than to $k_{z}\mathrm{=}0$. 
 
For the undoped $\left( d^{9}\right) $ cuprates, the LDA bandstructures are 
roughly similar to this, but the antibonding $3z^{2}-1$ band is now full and 
lies in the region of the $t_{2g}$ bands$.$ Filling this band has 
annihilated the $pd\sigma $ bond to apical oxygen and thereby caused $z_{%
\mathrm{Cu}\text{-}\mathrm{O}}$ to increase well beyond $x_{\mathrm{Cu}\text{%
-}\mathrm{O}}$, whereby the antibonding push-up of the $3z^{2}-1$ band has 
been lost. The half-full $pd\sigma $ antibonding $x^{2}-y^{2}$ band lies a 
bit lower with respect to the O and cation bands than in the nickelates 
because the position of the $3d$-level in Cu is lower than in Ni. However, 
the shape of this cuprate conduction band near half filling is not unlike 
that of the lowest $e_{g}$ band in the nickelate heterostructures, in 
particular for the cuprates with the highest $T_{c\,\max }$. Specifically, 
LDA calculations for a large number of cuprate families have revealed that 
whereas the dispersion along the nodal direction (ZA) is always the same, 
the energy of the saddlepoints at $\left( \frac{\pi }{a},0\right) $ and $%
\left( 0,\frac{\pi }{a}\right) ,$ i.e. at R, depends on the material and is 
lower for materials with higher $T_{c\,\max }$\cite{Pavarini01}. The reason 
for this correlation is not understood, but the reason for the change of 
band shape is clearly that the $x^{2}-y^{2}$ orbital is hybridizing with a 
material-dependent \emph{axial} $\left( \left\vert m\right\vert \mathrm{=}%
0\right) $ \emph{orbital} whose energy lies $\sim $10 eV above the Fermi 
level, but falls for cuprates with increasing $T_{c\,\max }$. This axial 
orbital is essentially the antibonding linear combination of Cu $4s$ and 
apical O $2p_{z},$ so that its energy falls if their interaction decreases, 
e.g. by increasing $z_{\mathrm{Cu}\text{-}\mathrm{O}}$. Concomitant with 
this change of band shape is a concentration of the conduction-band Wannier 
function onto the CuO$_{2}$ layer, away from the perpendicular direction. 
Instead of using the energy of the axial orbital as band-shape parameter, 
one uses a dimensionless parameter, $r$, which for materials with low $%
T_{c\,\max }$ $\left( <50K\right) $ becomes the ratio $t^{\prime }/t$ of the 
2nd to the 1st-nearest-neighbor hopping integral. The cuprates with the 
highest $T_{c\,\max }$ $\left( \sim 140\,\mathrm{K}\right) $ have $r\sim 0.4. 
$ If one could lower the energy of the axial orbital right down to the Fermi 
level, $r$ would have the value $1/2$. 
 
This axial-orbital model also applies to the $e_{g}$ bands of nickelate 
heterostructures, but due to the short distance to apical oxygen, the axial 
orbital is now essentially the antibonding linear combination of Ni $%
3d_{3z^{2}-1}$ and apical O $2p_{z}.$ Its energy is that of the $3z^{2}-1$ 
band at $\Gamma ,$ and since this is \emph{below} $\varepsilon _{F}$ for the 
LDA bands shown in Fig. 1, they have $r>1/2.$ Engineering these 
heterostructures should presumably first aim at reducing $r$ towards that $%
\left( \mathrm{\sim }0.4\right) $ of the cuprates with the highest $%
T_{c\,\max },$ i.e. at moving the energy of the second band at $\Gamma $ 
well above $\varepsilon _{F}.$ This requires increasing the interaction 
between Ni $3d_{3z^{2}-1}$ and apical O $2p_{z},$ e.g. by reducing $z_{%
\mathrm{Ni}\text{-}\mathrm{O}}$. As we shall see, this is helped by the 
electronic correlations, but does not necessarily lead to HTSC, because 
although the same value of $r$ gives the same band shape for nickelates and 
cuprates, their conduction-band Wannier orbitals are not identical. 
 
Having studied the materials dependence of the LDA band structures, and 
having found that the conduction bands in the paramagnetic phase are well 
separated from all other bands, we can study the effects of Coulomb 
correlations in the nickelate heterostructures using the two-band Hubbard 
Hamiltonian:%
\begin{eqnarray*} 
\hat{H} &=&\sum_{\mathbf{k,}mm^{\prime },\mathbf{\sigma }}H_{mm^{\prime }}^{%
\mathbf{k}}\,\hat{c}_{m\sigma }^{\mathbf{k}\,\dagger }\,\hat{c}_{m^{\prime 
}\sigma }^{\mathbf{k}}\;+\;U\sum_{i,m}\hat{n}_{m\uparrow }^{i}\,\hat{n}%
_{m\downarrow }^{i} \\ 
&&+\sum_{i,mm^{\prime },\sigma \sigma ^{\prime }}\left( V-\delta _{\sigma 
\sigma ^{\prime }}J\right) \hat{n}_{m\sigma }^{i}\,\hat{n}_{m^{\prime }\neq 
m\,\sigma ^{\prime }}^{i}\,. 
\end{eqnarray*}%
Here, the on-site Coulomb terms, namely the intra and inter-orbital Coulomb 
repulsions, $U$ and $V=U-2J,$ as well as the Hund's exchange, $J,$ have been 
added to the LDA $e_{g}$ Wannier-function Hamiltonian, $H_{mm^{\prime }}^{%
\mathbf{k}}.$ This Hubbard Hamiltonian we solve for 1/4 filling in the 
single-site DMFT approximation for the paramagnetic phase and at a 
temperature so high $\left( 1160\,\mathrm{K}=0.1\,\mathrm{eV/k}_{\mathrm{B}%
}\right) $ that we can afford using the Hirsch-Fye Quantum Monte Carlo 
method. 
 
\begin{figure}[tbh] 
\centering   \includegraphics[width=4.0cm]{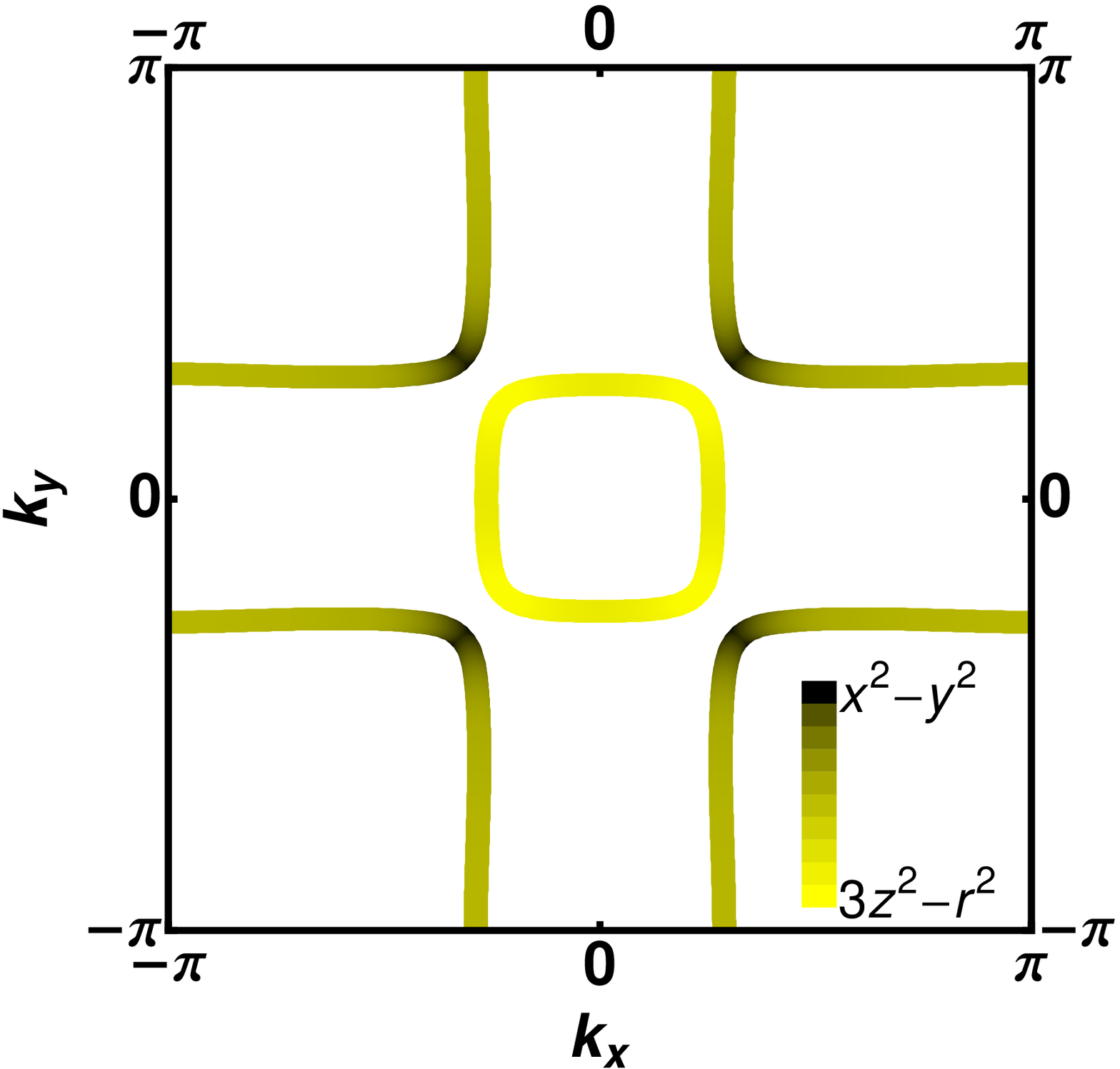}  
\includegraphics[width=4.0cm]{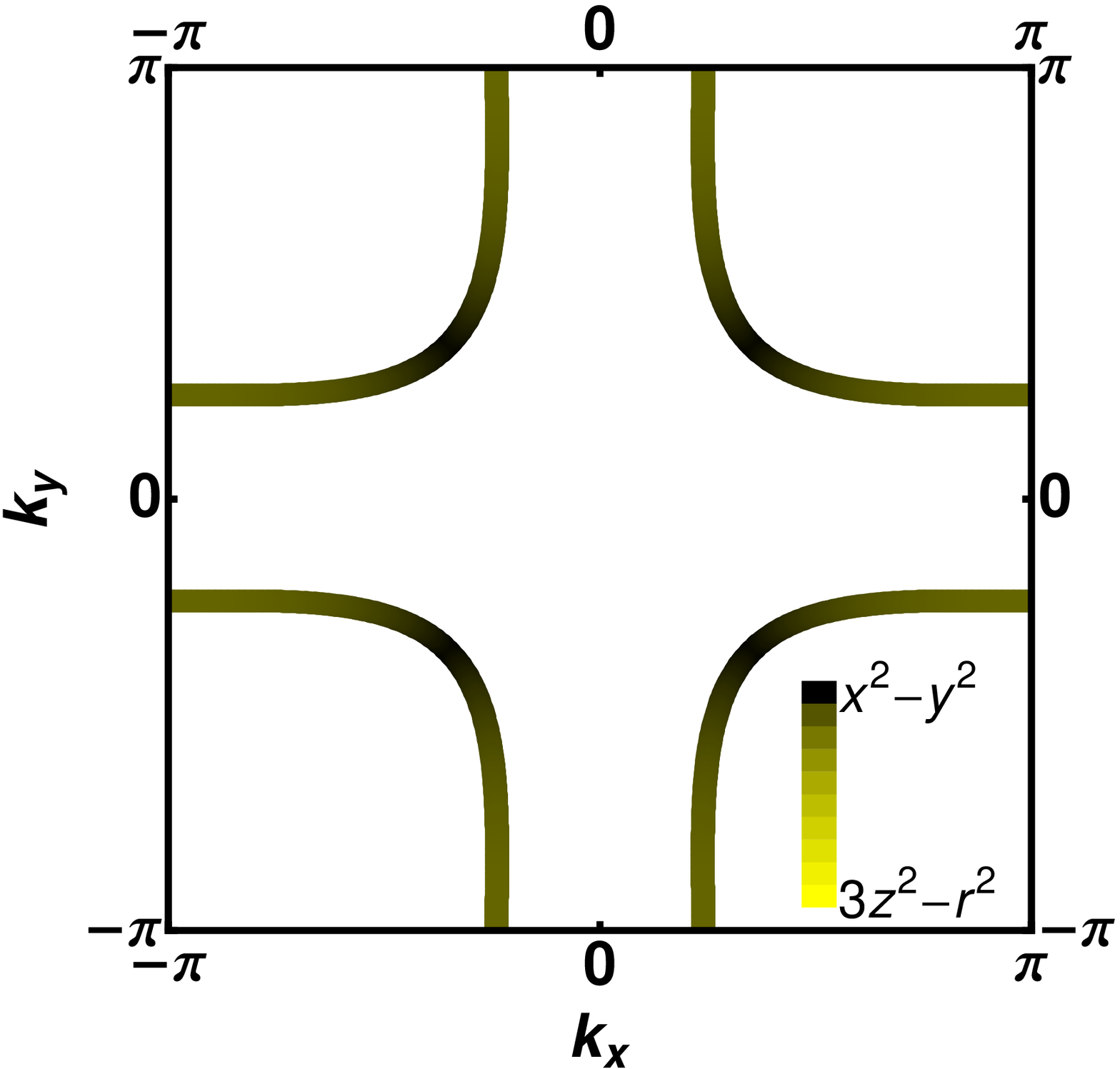} 
\caption{Cross-section of the LDA (left) and LDA+DMFT ($U=6.7$ eV) FS with 
the $k_{z}\mathrm{=}0$ plane for the unstrained 1/1 heterostructure. $e_{g}$-orbital characters are coded as in Fig.$\,$1.} 
\label{figure 2} 
\end{figure}

Our DMFT calculations confirm the 
common expectation that, for a metallic multiband system, the main effect of 
the Coulomb correlations is to enhance the splitting between the subbands 
such as to reduce the density of states at the Fermi level. Specifically, 
for the undoped superlattice with $J\mathrm{=}0.7\,$eV and $U$ increasing, 
we find that the bottom of the $3z^{2}-1$ band is driven up and passes the 
Fermi level when $U$ exceeds 6.4 eV for the unstrained and 5.7 eV for the 
strained structure. Hereafter the FS has only one sheet, a large $\left(  
\frac{\pi }{a},\frac{\pi }{a}\right) $-centered hole cylinder whose shape 
can be seen from Fig. 2 to be similar to that found in the cuprates with the 
highest $T_{c\,\max },$ but even more extreme. It is of course possible that 
the strong nesting of this FS makes it unstable with respect to spin or/and 
charge-density waves with $q_{x}\mathrm{\sim }\frac{\pi }{2a}$ and $q_{y}%
\mathrm{\sim }\frac{\pi }{2a},$ similar to what has been found in cuprates. 
At the point where the second sheet disappears, $r\mathrm{=}1/2$ and the 
ratio between the $x^{2}-y^{2}$ and $3z^{2}-1$ populations has increased to 
80/20 for the unstrained -- and beyond for the strained -- superlattice. 
Reasonable changes of $J$ slightly influence details of the Hubbard 
subbands, but not the physics of the transition. 
 
The remaining half-full band undergoes a Mott-transition when $U$ exceeds 
7.4 eV for the unstrained and $6.5$ eV for the strained superlattice. For 
comparison, a half-full cuprate band undergoes a Mott transition in DMFT for 
a critical value of $U$ which increases with $r$ and takes the value $4.5$ 
eV for $r\mathrm{=}0.4$ \cite{Saha}. This behavior for the cuprates is thus 
in line with what we find for the nickelate heterostructures where $r_{%
\mathrm{LDA}}\left( \mathrm{unstrained}\right) $ $>$ $r_{\mathrm{LDA}}\left(  
\mathrm{strained}\right) $ $\sim $ $1/2,$ and this supports our hope that 
the nickelates can be engineered such that, like in the cuprates, 
hole-doping will suppress the Mott transition and produce superconductivity. 
For nickelates there is even the possibility of engineering the $e_{g}$ 
bands such that the real value of $U$ falls between the one needed to reduce 
the FS to a single sheet and the one needed to eliminate this sheet by a 
Mott transition. If this can be achieved, superconductivity in the 
nickelates may occur even without doping. This is a remarkable result.

\begin{figure}[tbh] 
\includegraphics[width=5.0cm]{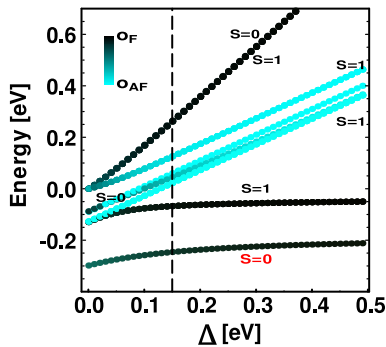} 
\caption{Energy levels for the unstrained two-site model with $U=6.4$ eV as 
a function of the splitting $\Delta $ between the energies of the $3z^{2}-1$ 
and $x^{2}-y^{2}$ Wannier orbitals. The LDA value of $\Delta $ is indicated 
by the dashed line. $O_{F}$ ($O_{AF}$) denotes a configuration with the same 
(different) orbital(s) on the two sites.} 
\label{Fig:levelplot} 
\end{figure} 
 
Next, we need to estimate the strength of antiferromagnetic correlations, 
which are believed to play a central role in the physics of the cuprates. 
Since our LDA+DMFT calculations would be prohibitively expensive for the 
study of low temperature magnetic properties, we merely diagonalized the 
two-site version of the Hubbard Hamiltonian obtained by Fourier 
transformation of $H_{mm^{\prime }}^{\mathbf{k}}$ and truncation to a 
diatomic molecule directed along $x.$ The energy levels are presented in Fig.%
$\,$\ref{Fig:levelplot} as functions of the the difference, $\Delta ,$ 
between the energies of the $3z^{2}-1$ and $x^{2}-y^{2}$ Wannier orbitals. 
The ground state is always a spin singlet. Increasing $\Delta $ from $0$ to $%
\infty $ leads to demixing such that the orbital configuration changes from $%
3x^{2}-1$ to $x^{2}-y^{2}.$ For the LDA value, the orbital character is 
already close to $x^{2}-y^{2}.$ From the distance between the singlet ground 
state and the triplet first excited state, we estimate the magnitude of the 
antiferromagnetic coupling constant to be $J_{AF}\sim 0.2\,$eV, i.e. 
somewhat higher than in cuprates. 
 
Altogether, our analysis of the 1/1 LaNiO$_{3}$/LaAlO$_{3}$ system shows 
that heterostructuring of $d^{7}$ nickelates is promising because their 
physics contains the main ingredients of high-temperature superconductivity. 
In particular, we find that electronic correlations reduce the FS to a \emph{%
single} sheet whose shape is similar to the one in the hole-doped cuprates 
with the highest $T_{c\,\text{max}}.$ This sheet has not only $x^{2}-y^{2},$ 
but also $3z^{2}-r^{2}$ character, and this gives a new twist to the 
intensive discussion of orbital-selective Mott-Hubbard transitions. 
Substrate-induced strain and/or use of insulating layers different than LaAlO%
$_{3}$ may tune the FS shape and may enable superconductivity without 
doping.  
 
Discussions with J. Chakhalian, H.-U. Habermeier, and T. Saha-Dasgupta are 
gratefully ackowledged.


\begin{references}   
   
\bibitem{Bednorz86}   
J.G. Bednorz and K.A. M\"uller, Z.\ Phys B {\bf 64}, 189 (1986).   
   
\bibitem{Maeno94}   
Y. Maeno \textit{et al.}, Nature {\bf 372}, 532 (1994). 
   
\bibitem{Takada03a}   
K. Takada  \textit{et al.}, Nature {\bf 422}, 53 (2003). 
   
   
   
\bibitem{Chaloupka}   
J. Chaloupka and G. Khaliullin, Phys.\ Rev.\ Lett. {\bf 100}, 016404 (2008).   
   
\bibitem{Anisimov99}   
 V.I. Anisimov, D. Bukhvalov, and T.M. Rice, Phys. Rev. B {\bf 59}, 7901 (1999).   
   
 
\bibitem {Chakh07} J. Chakhalian \textit{et al.}, 
Nat. Phys. {\bf 2}, 244 (2006); J. Chakhalian \textit{et al.}, Science {\bf 318}, 1114 (2007). 
   
 
\bibitem{OSMT} See, e.g., V.I. Anisimov {\sl et al.}, Euro. Phys. J. B {\bf 25}, 191 (2002);   
A. Koga \textit{ et al.}, 
Phys. Rev. Lett. {\bf 92}, 216402 (2004); R. Arita and K. Held, Phys. Rev. B   
{\bf 72}, 201102(R) (2005); A.I. Poteryaev {\sl et al.}, Phys. Rev. B {\bf 76}, 085127 (2007).  
 
\bibitem{LDADMFT} 
V.I. Anisimov \textit{ et al.}, J. Phys.: Condens. Matter {\bf 9}, 7359 (1997);   
A.I. Lichtenstein and M.I. Katsnelson, \newblock Phys. Rev. B {\bf 57}, 6884 (1998); 
G. Kotliar \textit{ et al.},  Rev. Mod. Phys.  {\bf 78}, 865  (2006). 
   
   
\bibitem{DMFT}   
W. Metzner and D. Vollhardt, \newblock Phys. Rev. Lett. {\bf 62},  324  (1989); A. Georges \textit{ et al.}, 
Rev. Mod. Phys. {\bf 68},  13 (1996). 
   
\bibitem {Pavarini01} E. Pavarini \textit{ et al.}, Phys. Rev. Lett. \textbf{87}, 047003 (2001). 
  
\bibitem {VASP} G. Kresse, J. Furhmuller, Software VASP, Vienna (1999); 
P.E. Bl\"ochl, Phys. Rev. B {\bf 50}, 17953 (1994). 
 
   
\bibitem {NMTO} O.K. Andersen and T. Saha-Dasgupta,   
Phys. Rev. B {\bf 62}, R16219 (2000). 
   
\bibitem{ttprime}  O.K. Andersen  \textit{ et al.}, J. Phys. Chem. Solids {\bf 56}, 1573 (1995).   
 
  
   
   
   
 
\bibitem{Saha} T. Saha-Dasgupta (unpublished); see also H. Das and T. Saha-Dasgupta, 
Phys. Rev. B (accepted). 
   
   
\end{references}
\end{document}